\documentclass[conference]{IEEEtran}
\usepackage[hyphens]{url}
\usepackage{hyperref}
\hypersetup{breaklinks=true}
\hypersetup{draft}

\usepackage{comment}
\usepackage{enumitem}
\usepackage{multirow}
\usepackage{float}

\ifCLASSINFOpdf
  \usepackage[pdftex]{graphicx}
  \graphicspath{{../pics/}}
  \DeclareGraphicsExtensions{.pdf,.jpeg,.png}
\else
\fi
\usepackage{tikz,ifthen}
\usepackage{verbatim}

\usepackage{amsmath}
\usepackage{algorithm}
\usepackage{algorithmic}

\usepackage[utf8]{inputenc}
\usepackage{fourier} 
\usepackage{array}
\usepackage{makecell}

\begin{document}
	\bstctlcite{IEEEexample:BSTcontrol}
	
	\raggedbottom
	
%
\title{Multiple Patients Behavior Detection in Real-time\\ using mmWave Radar and Deep CNNs}
%
%
%

\author{
	\IEEEauthorblockN{Feng Jin, Renyuan Zhang, Arindam Sengupta, \\ Siyang Cao and Salim Hariri}
	
	\IEEEauthorblockA{Department of Electrical and Computer Engineering\\
		The University of Arizona, Tucson, AZ 85721\\
		Email:\{fengjin, ryzhang, sengupta, caos\}@email.arizona.edu \\ and hariri@ece.arizona.edu}
	
	\and
	
	\IEEEauthorblockN{Nimit K. Agarwal and Sumit K. Agarwal}
	
	\IEEEauthorblockA{Department of Medicine\\
		Banner - University Medical Center Phoenix\\
		Phoenix, AZ 85006\\
		Email:\{Nimit.Agarwal, Sumit.Agarwal\}\\@bannerhealth.com}
}	
	
\maketitle

\begin{abstract}
To address potential gaps noted in patient monitoring in the hospital, a novel patient behavior detection system using mmWave radar and deep convolution neural network (CNN), which supports the simultaneous recognition of multiple patients’ behaviors in real-time, is proposed. In this study, we use an mmWave radar to track multiple patients and detect the scattering point cloud of each one. For each patient, the Doppler pattern of the point cloud over a time period is collected as the behavior signature. A three-layer CNN model is created to classify the behavior for each patient. The tracking and point clouds detection algorithm was also implemented on an mmWave radar hardware platform with an embedded graphics processing unit (GPU) board to collect Doppler pattern and run the CNN model. A training dataset of six types of behavior were collected, over a long duration, to train the model using Adam optimizer with an objective to minimize cross-entropy loss function. Lastly, the system was tested for real-time operation and obtained a very good inference accuracy when predicting each patient’s behavior in a two-patient scenario.
\end{abstract}

\begin{IEEEkeywords}
Behavior detection, fall detection, mmWave radar, Doppler pattern, CNN. 
\end{IEEEkeywords}

%
\IEEEpeerreviewmaketitle

\section{Introduction}
\par Continued direct observation for hospitalized patients or residents in post-acute care settings, especially those who are cognitively impaired from various causes (e.g. Alzheimer’s dementia), is difficult and largely deficient due to limited manpower and nursing resources. This can lead to windows of "unsupervised care" which can further lead to serious safety concerns, such as inadvertent falls, missed emergencies like detecting seizures, detecting early signs of delirium, agitation, etc. This motivates researchers and engineers to come up with an automated solution to detect and report these untoward patients behavior during the period of "unsupervised care".

\par The basic idea of patient behavior detection is to use various sensors to collect data for different kinds of motion, and then apply a classifier to recognize the behavior. K. Chaccour et al. in \cite{ref_MOTION_DETECTION_SYS} divided the detection systems into three categories, viz. wearable based, non-wearable based and fusion based systems. Wearable based approach is to have a patient carry on lightweight devices integrated with inertial and/or magnetic sensors for data collection. A. T. Ozdemir et al. in \cite{ref_WEARABLE_ACC2} used multiple Inertial Measurement Units (IMUs) to extract total acceleration vector as behavior features and applied a k-nearest neighbor (k-NN) algorithm to detect fall. J. Gubbi et al. in \cite{ref_WEARABLE_SEIZURE} developed a wrist-worn device integrated with an accelerometer and applied a support vector machines (SVMs) method to determine if the patient has epileptic seizure (ES) or psychogenic nonepileptic seizure (PNES) or both. On the other hand, non-wearable approach does not use body-fixed sensors. Instead, sensors like color-based camera, Wi-Fi, RGB depth (RGB-D) camera, acoustic sensor, infrared sensor, are deployed in the environment. With color-based camera, H. Lu et al. in \cite{ref_CAM_SEIZURE} used segmentation to track the movement of limbs, and detect epileptic seizure based on displacement and oscillation features of limbs. In \cite{ref_WIFI_FALL}, Y. Wang et al. proposed a WiFall system, which utilized the channel status information (CSI) in Wi-Fi signals as an indicator for different motions, and then used SVMs to classify them into resting, falling, sitting down, walking and standing up. In \cite{ref_RGBD_FALL2}, doctors presented experimental finds to show RGB-D sensor's potential benefits for fall detection. By using a commercially available RGB-D camera, the Microsoft Kinect v2, A. Amini et al. in \cite{ref_RGBD_FALL} implemented both heuristic based and machine learning based algorithms to detect fall.

\begin{figure*}[ht]
	\centering
	\includegraphics[width=7in]{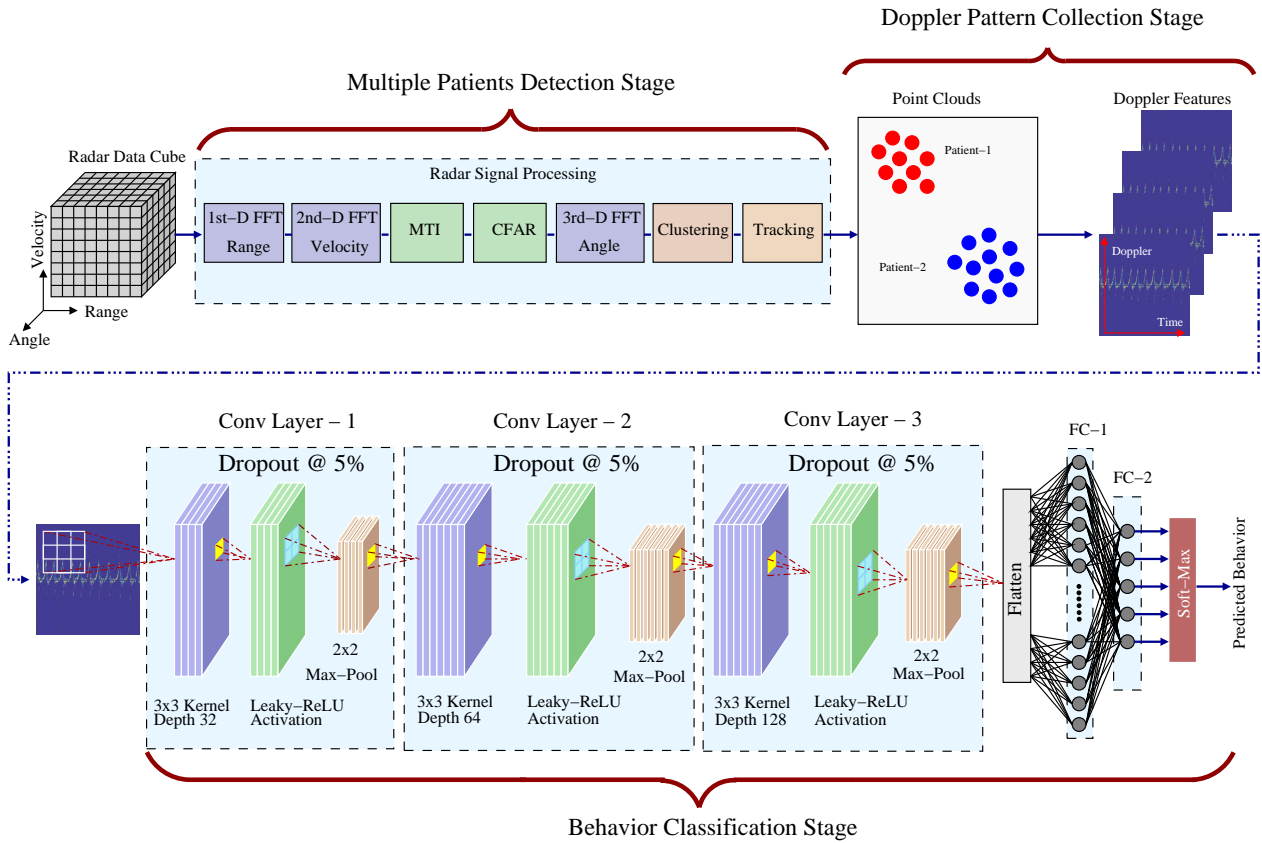}
	\caption{An overview of the proposed architecture: The acquired raw radar data is first processed to identify and localize multiple patients in the environment. The Doppler patterns for each of the clusters are then passed through a Deep CNN followed by two fully-connected (FC) layers that output predicted behavior for each localized patient.}
	\label{fig_ProposedSystem}
\end{figure*}

\par There has been emerging interest among researchers to detect human behavior using radar sensors. As one of the non-wearable methods, radar sensor does not require patients' compliance to wear or make them feel uncomfortable. Furthermore, radar sensor has attractive advantages over camera based systems in terms of privacy. Most importantly, apart from distance, radar sensor can measure the velocity directly and precisely which is essentially valuable to motion detection, compared to RGB-D sensor which can only measure velocity indirectly by differentiating two consecutive frames. B. Jokanovic et al. in \cite{ref_RADAR_FALL1}\cite{ref_RADAR_FALL2} used a vector network analyzer (VNA) working as a continuous wave (CW) radar, with carrier frequency at 6GHz, to collect Doppler patterns of test subjects, and then apply a neural network to classify the motions. Shengheng Liu et al. in \cite{ref_RADAR_FALL3} applied short-time fractional Fourier transform (STFrFT) on the data collected by a C-band frequency-modulated continuous-wave (FMCW) radar, and then detected fall using a Bayesian classifier. Sevgi Z. Gurbuz et al. had done a similar research but with convolutional neural network (CNN) as the classifier \cite{ref_RadarCNN1} \cite{ref_RadarCNN2}. However, until 2017, just one year prior to when our research was conducted \cite{ref_prevWork}, none of them used a higher frequency radar operating in W band, for example in 77GHz or 90GHz, as highlighted in \cite{ref_RADAR_REVIEW}. Also, these radar sensor based researches did not show the ability to detect multiple patients' behavior, simultaneously.

\par In this paper, we propose a multiple patients behavior detection system using millimeter wave (mmWave) radar sensor operating at 77GHz. The most attractive advantage of mmWave radar is its high resolution due to high bandwidth (up to 4GHz), and is also inexpensive and small in size. In this proposed system, we first use the mmWave radar sensor to track each patient in a ward and collect the Doppler pattern of his/her torso and limbs over a period of time. And then a deep CNN is created to classify his/her behavior, like walking, falling, swing hand for help, seizure, restless movement, etc. 

\par In Section II, we describe the proposed system in detail, including the radar signal processing algorithm we used to track and detect multiple patients, how we collected the Doppler pattern for each patient and the structure of the deep CNNs we created to classify the behavior. Section III describes the experimental setup in a conference room to emulate a ward in a hospital, the real-time inference accuracy when predicting two patients' behavior simultaneously is also presented. In section IV, we provide inferred conclusions and summarize several potential ways to improve the system.

\section{Proposed System using mmWave Radar and Deep CNN}
\par Due to the high resolution offered by the mmWave radar sensor, a human body would have several scattering points reflecting the incident mmWave signals. Primarily, they are due to, (i) the torso, that has a large reflection area, and (ii) the limbs (legs and arms). This leads to the formation of a point cloud on account of the many scattering points from different parts of the body. Meanwhile, these scattering points will have different velocities, i.e. Doppler shifts, due to the different type of motions of the individual body parts. As an example, the left arm will have a velocity in a direction opposite to right arm when people walk while swinging their arms. Similarly, different behavior would have various movement patterns of a person's torso and limbs, and the resulting Doppler pattern of the point cloud would thereby differ from each other.

\par In the proposed system, we use a fast chirp FMCW mmWave radar sensor to track multiple patients in a ward simultaneously and detect the point cloud for each patient. Then a period of Doppler pattern from each patient's limbs and torso are collected. Finally we use a deep CNN model to classify the behavior from each patient. Fig. \ref{fig_ProposedSystem} shows the schematic of proposed detection system.

\subsection{Multiple Patients Detection and Tracking using mmWave Radar}
\par The mmWave radar sensor sends out a FMCW chirp, with carrier frequency on 77GHz, and uses stretching processing to get the beat frequencies related to the range of scattering points. To solve for velocity, also in terms of Doppler shift, multiple chirps are sent out, and then the Doppler shift across chirps during each coherent processing interval (CPI) can be found by analyzing the data in frequency domain. Each mmWave radar sensor has multiple receiving antenna channels placed in azimuth, where beamforming method is used to solve for the angle of each scattering point. The collected raw data is formed in a three-dimensional datacube. Range and Doppler processing are then performed on the datacube first, followed by moving target indication (MTI) to remove the static clutter points, i.e. reflection from walls, desks, etc. To detect the scattering points from a noisy background, constant false alarm rate (CFAR) detection method is used, followed by the angle estimation for each detected point.
\par A clustering method like density-based spatial clustering of applications with noise (DBSCAN) \cite{ref_DBSCAN} is used to separate these scattering points into multiple targets, i.e. multiple patients in our case. Finally, a Kalman filter is applied to track each patient's trajectory and associate the point cloud of each patient with a trackID. 

\subsection{Doppler Pattern Collection}
\par There are several techniques to collect Doppler features over time, such as micro-Doppler analysis using short-time Fourier transform (STFT), or Wavelet transform. In our case, for simplicity, a period of Doppler bins of those scattering points with the same trackID are collected by using a sliding window. At the same time, the intensity of scattering point is compensated for the attenuation due to the range effect, and normalized to 1 before passing to the neural network for training, to ensure consistent bounds on intensity over the training data. The normalized Doppler pattern of the point cloud works as a behavior signature, that is used in the classification stage.

\subsection{Behavior Classification using Deep CNN}
\par To classify behaviors, we build a deep CNN model that consists of three CNN layers, each with a $3 \times 3$ kernel size, and depths 32, 64 and 128 respectively. Each neuron output is activated with a Leaky-Relu function, instead of a conventional Relu activation function, to overcome the `dying Relu' effect \cite{ref_DyingReLu} which potentially leads to certain neurons to permanently remain in an inactive state. To reduce computational complexity further, every CNN layer is followed by a 2-D max-pooling layer. The max-pooling layer downsamples the output from the CNN preceding it, while preserving the most dominant features detected from the previous stage. This is done by sliding a $2 \times 2$ window across the CNN output, and returning the maximum pixel value in its stride.

\par To avoid overfitting, we use dropout regularization between layers, with an individual node dropout probability of $5\%$. This means, in every training cycle some nodes would `drop-out' or detach itself from the computation graph, for both forward and backward propagation, and the training parameters would be optimized for the rest of the nodes. In the next cycle, these nodes would be re-inserted, and some other nodes would dropout, and the process would continue throughout the training phase. This also means that for a given epoch, we are now having to compute and optimize fewer parameters, thereby increasing the speed of training. The outputs from the final CNN layer is then flattened to a 1-D vector and is subject to the fully-connected (FC) layers. The final output layer would have $k$ nodes, corresponding to the $k$ classes of human behavior we aim to detect. The outputs are then normalized and associated with probabilities of each of the classes, using the softmax function. The class with the greatest associated probability would be the predicted human behavior.

\section{Experiment and Results}
To verify the effectiveness of the proposed system, we conducted several experiments to detect five critical behaviors, viz. walking, falling down to the floor, swing hand for help, seizure, restless movement. First, we implemented the proposed system on hardware. And then we collected and labelled the training datasets for these five behaviors, and some others to detect transition and no-activity. These datasets were used to train the deep CNN model. After that, we ran the model in real-time to find the inference performance. 

\subsection{Experiment Setup}
\begin{figure}[ht]
	\centering
	\includegraphics[width=3in]{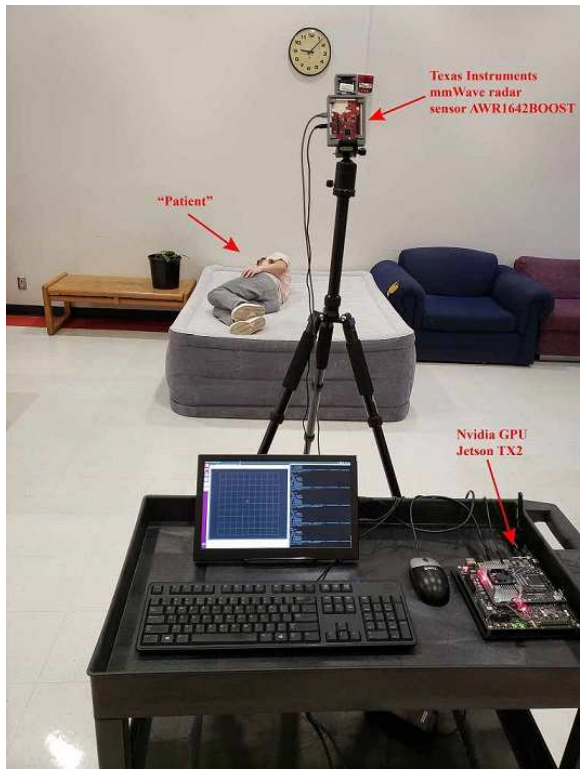}
	\caption{Experimental setup to emulate a hospital ward.}
	\label{fig_experimentSetup}
\end{figure}

We deployed the radar signal processing part in Fig. \ref{fig_ProposedSystem} into the Texas Instruments (TI) AWR1642BOOST mmWave radar platform \cite{ref_AWR1642}, by taking advantage of a demo project from TI, and developed the Doppler pattern collection and CNN-based behavior classification parts on the Nvidia Jetson TX2 platform. The mmWave radar sensor tracked each moving patient, detected its point cloud and then transferred the data to the Nvidia TX2 through a serial port. The mmWave radar parameters used in our setup are listed in Table \ref{tab_radarparam}. On the Ubuntu-running TX2 platform, robot operating system (ROS) was used to create three ROS nodes. The first one was to receive the streaming point cloud data from TI mmWave radar and pass it to the second node, which was to collect one second duration of Doppler pattern for each trackID, i.e. each patient. The third node invoked the deep CNN model to classify behaviors based on the Doppler patterns from network input. The experiment setup is shown in Fig. \ref{fig_experimentSetup}. The experiments were conducted in a conference room in a similar setting to a hospital ward, and the author in this paper acted as a patient for testing.
\begin{table}[h]
	\renewcommand{\arraystretch}{1.3}
	\caption{mmWave Radar Waveform Configuration in Experiment}
	\label{tab_radarparam}
	\centering
	\begin{tabular}{|c||c||c|}
		\hline
		\bfseries Parameter & \bfseries Value & \bfseries Unit \\
		\hline
		Start Frequency & 77 & GHz \\
		\hline
		Bandwidth & 3.072 & GHz\\
		\hline
	 Chirp Rate & 60 & MHz/us \\
	 	\hline
	 	ADC Sampling Rate & 2.5 & MHz\\
		\hline
		Samples per chirp & 128 & \\
		\hline
		Chirps per frame & 256 &\\
		\hline
		Frame duration & 50 & ms \\
		\hline
		Range resolution & 0.0488 & m\\
		\hline
		Max unambiguous range & 5 & m \\
		\hline
		Velocity resolution & 0.0827 & m/s\\
		\hline
		Max radial velocity & 5.2936 & m/s\\
		\hline
		Azimuth angle resolution & 14.5 & deg\\
		\hline
	\end{tabular}
\end{table}

\subsection{Data Collection and Labelling}
\par
\begin{figure*}[ht]
	\centering
	\includegraphics[width=7in]{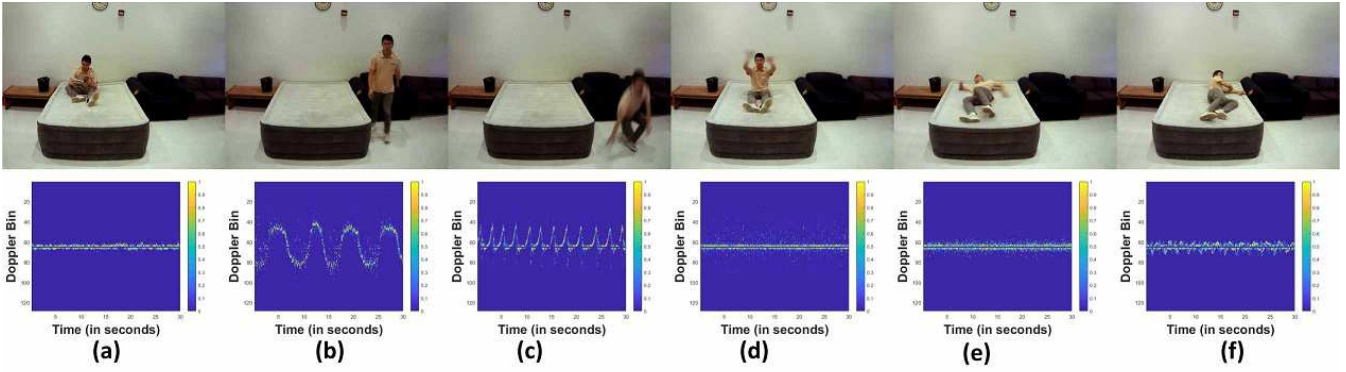}
	\caption{Doppler pattern for each behavior with associated experiment scenario. (a) Other behavior. (b) Walking. (c) Falling. (d) Swing hand for help. (e) Seizure. (f) Restless movement.}
	\label{fig_DopplerPattern}
\end{figure*}

 We collected the training dataset with only one patient. The patient continuously performed each one of the behaviors listed in the Table \ref{tab_traingdataset}, and the data samples were collected in a ROS bag format, where each sample was the Doppler pattern of specified behavior with one second duration. We labelled these samples with a integer number each corresponding to a distinct human behavior. As the duration of falling is too short, in an alternate way, we collected the data with a stand-fall-stand repetition, and then manually truncated the Doppler patterns to obtain the clean data for falling only. As in this study we primarily focus on detecting walking, falling, swing, seizure and restless movement, all other behaviors is labelled with 0. The other behaviors cover all the other situations, such as sitting with eating food, reaching hand to take something, etc. If we did not cover a situation, say, standing while stretching arms, this might lead to being incorrectly classified into one of the 5 aforementioned classes of interest. Therefore, it needs a lot of work to collect the dataset for various other behaviors. For simplicity, in our case, except for the five critical behaviors, we collected several other behaviors like standing with watching smartphone and sitting with reading a book, etc. Finally, we obtained the Doppler pattern over 30 seconds for each behavior in the experimental setting, as shown in Fig. \ref{fig_DopplerPattern}.
\begin{table}[H]
	\renewcommand{\arraystretch}{1.3}
	\caption{Training Dataset}
	\label{tab_traingdataset}
	\centering
	\begin{tabular}{|c||c||c|}
		\hline
		\bfseries Behavior & \bfseries Samples & \bfseries Label Value \\
		\hline
		Other & 41,788 & 0\\
		\hline
		Walking & 11,201 & 1\\
		\hline
		Falling & 5,745 & 2\\
		\hline
		Swing & 10,719 & 3\\
		\hline
		Seizure & 10,299 & 4\\
		\hline
		Restless Movement & 17,216 & 5\\
		\hline
	\end{tabular}
\end{table}

\subsection{Training}
\par To train the model, we used Keras \cite{ref_Keras} application program interface (API) to build the deep CNN model that works on a TensorFlow \cite{ref_tensorflow} framework. The Adam \cite{ref_adam} optimizer with mini-batch of 64 was chosen to update the model parameters to approach the minimum loss under cross-entropy criteria. The validation dataset was a randomly selected 10\% from the entire dataset. We trained the model on a desktop computer employing an Nvidia GeForce 1050 GPU for 10 epochs. As the number of epochs increased, the loss was decreased and the validation accuracy was increased. This indicated that the loss function was converging to the minimum value and that the dataset we collected was valid. Upon training completion, the final loss was 0.0365 and the test accuracy was 98.69\%. 

\par To evaluate how the size, diversity of the dataset and the depth of CNN model effected the loss and accuracy, we trained the model in several different configurations listed in Table \ref{tab_traingconfig}. Firstly, we started from 3 layer CNNs with trainable parameters of 880,006. Comparing configuration \#1 with \#2, we reduced the dataset size to 70\% of the entire dataset, then the loss was increased by 27\% and the accuracy was decreased, which means more data helps to improve the model accuracy. Secondly, we increased the CNN layers from 3 to 4, making the model deeper with 1,438,214 parameters. Then comparing configuration \#1 with \#3, the loss increased, however the accuracy increased as well. This indicated overfitting, and therefore, a deeper model did not help to reduce the loss for the given dataset. Thirdly, we retained the dataset for the five critical behaviors, but diversified the dataset of other behaviors. What we did was collect almost the same length dataset for other behaviors, but with more variety of behaviors like standing with stretching the body, sitting with moving the chair back and forth a little bit, etc. Then comparing configuration \#1 with \#4, the loss increased by 77\% and the accuracy dropped. This is because more diverse dataset increased the complexity of the loss function, indicating that it would require deeper layers and more epochs to close to the minimum.
\begin{table}[H]
	\renewcommand{\arraystretch}{1}
	\caption{Different Training Configuration}
	\label{tab_traingconfig}
	\centering
	\scalebox{0.85}{\begin{tabular}{|c||c||c||c||c||c||c|}
		\hline
		\bfseries \# & \bfseries \makecell{CNN\\Layers}  & \bfseries \makecell{Model\\Parameters} & \bfseries \makecell{Training\\Dataset} & \bfseries \makecell{Validation\\Dataset} & \bfseries \makecell{Training\\Loss} & \bfseries \makecell{Test\\Accuracy}\\
		\hline
		1 & \makecell{3} & 880,006 & 90\% & 10\% & 0.0365 & 98.69\%\\
		\hline
		2 & \makecell{3} & 880,006 & 70\% & 30\% & 0.0464 & 98.09\%\\
		\hline
		3 & \makecell{4} & 1,438,214 & 90\% & 10\% & 0.0371 & 98.94\%\\
		\hline
		4 & \makecell{3} & 880,006 & 90\% & 10\% & 0.0646 & 97.77\%\\
		\hline
	\end{tabular}}
\end{table}

\subsection{Inference}
\par
\begin{figure}[ht]
	\centering
	\includegraphics[width=3.5in]{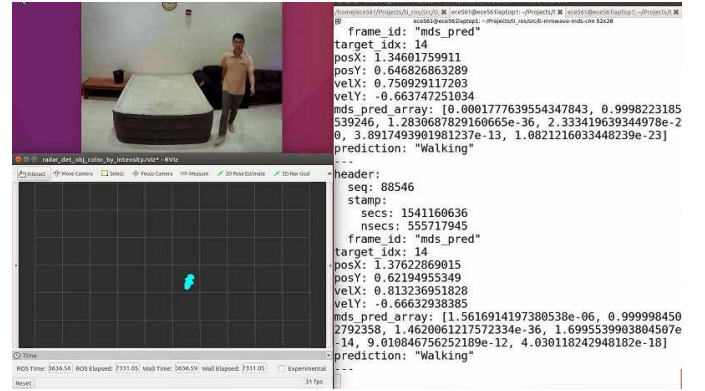}
	\caption{Real-Time behavior detection for a single patient scenario. The detected patient point cloud (in blue) and the deep CNN prediction ('walking') is shown.}
	\label{fig_OnePatient}
\end{figure}

 In the inference stage, we first conducted experiments to evaluate only one patient's behavior in real-time. As shown in Fig. \ref{fig_OnePatient}, the point clouds from the testing patient were collected and displayed on the ROS platform while the terminal output this patient's target id, position, velocity, and predicted patient's behavior as well. All the behaviors listed in the Table \ref{tab_traingdataset} were repeated for inference accuracy. We recorded behavior over a period of time, and obtained a series of prediction results over the same course. The inference accuracy was calculated by dividing the True-Positives (correct predictions) over the total number of predictions made. The inference accuracy results listed in Table \ref{tab_OnePatientPred} look promising for real case application. However, it needs to be noted that if we made the motion a little different from that when we were collecting for the training dataset, for example falling in another direction, the inference accuracy for falling would be downgraded. To overcome this, the training dataset should be collected to cover all kinds of situation.
\begin{table}[H]
	\renewcommand{\arraystretch}{1}
	\caption{Behavior prediction of one patient}
	\label{tab_OnePatientPred}
	\centering
	\begin{tabular}{|c||c||c|}
		\hline
		\bfseries Behavior & \bfseries Samples & \bfseries Inference Accuracy\\
		\hline
		Other & 1,337 & 95.74\%\\
		\hline
		Walking & 1,550 & 94.13\%\\
		\hline
		Falling & 1,560 & 84.49\%\\
		\hline
		Swing & 1,521 & 82.77\%\\
		\hline
		Seizure & 1,254 & 86.36\%\\
		\hline
		Restless Movement & 1,233 & 84.31\%\\
		\hline
	\end{tabular}
\end{table}

\par
\begin{figure}[ht]
	\centering
	\includegraphics[width=3.5in]{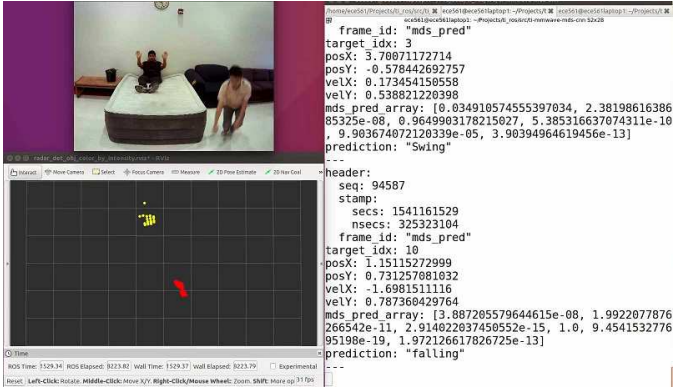}
	\caption{Real-Time behavior detection for a two patients scenario. The detected patients point clouds (in yellow and red) and the deep CNN prediction ('swing' for patient-1 and 'falling' for patient-2) is shown.}
	\label{fig_TwoPatient}
\end{figure}

 We then conducted experiments to calculate the inference accuracy for simultaneously detecting two patients' behavior. Fig. \ref{fig_TwoPatient} shows a two-patient scenario, where one was falling and the other one was swinging hand. The point clouds from these two patients were discriminated in different color for displaying, and the terminal could output the two different behaviors simultaneously in real-time. The predication results were collected for each target trackID, i.e. each patient, and then the inference accuracy was calculated for three different situations listed in Table \ref{tab_TwoPatientPred}.
\begin{table}[H]
	\renewcommand{\arraystretch}{1}
	\caption{Behavior prediction of two patient simultaneously}
	\label{tab_TwoPatientPred}
	\centering
	\begin{tabular}{|c||c||c||c||c|}
		\hline
			\multicolumn{2}{|c||}{\bfseries{Behavior}} & \multirow{2}{*}{\bfseries{Samples}} & \multicolumn{2}{c|}{\bfseries{Inference Accuracy}} \\
			\cline{1-2}   \cline{4-5}
			Behavior1 & Behavior2 & & Behavior1 & Behavior2\\
		\hline
		Walking & swing & 2,524 & 84.78\% & 79.11\%\\
		\hline
		Seizure & swing & 2,160 & 81.37\% & 88.41\%\\
		\hline
		Swing & falling & 2,230 & 85.24\% & 66.02\%\\
		\hline
	\end{tabular}
\end{table}

\section{Conclusion}
\par In this study, we used a mmWave radar to track and detect multiple patients, and created a deep CNN model to predict each patient's behavior. We collected dataset for six different kinds of behaviors, and trained a three-layer deep CNN model with very low loss and good test accuracy. Experiments were conducted to infer accuracy in a single patient and two patients scenarios with promising accuracy results. Based on the results, we conclude that a larger dataset would yield a better training loss, while, a deeper CNN may lead to overfitting. Furthermore, with more complex training data, we may need to train the model for more epochs. Finally, it is noted that the key to improve inference accuracy is to collect more data for varied motions under a variety of situations.
\ifCLASSOPTIONcaptionsoff
  \newpage
\fi

\bibliographystyle{IEEEtran}
\bibliography{./Multiple_Patients_Behavior_Detection_in_Real-time_using_mmWave_Radar_and_Deep_CNNs}

\end{document}